\documentclass[article,aps,showpacs,amssymb]{revtex4}

\usepackage{graphicx}% Include figure files

\begin{document}

\title{Poisson equation for weak gravitational lensing}

\author{Thomas P. Kling} \email{tkling@bridgew.edu}

\author{Bryan Campbell} \altaffiliation[Sponsored by]{~the Adrian
Tinsley Program for Undergraduate Research, BSC}

\affiliation{Dept. of Physics, Bridgewater State College,
Bridgewater, MA 02325}

\date{\today}

\begin{abstract}

\noindent Using the \citet{NP} spin coefficient (NP) formalism, we
examine the full Bianchi identities of general relativity in the
context of gravitational lensing, where the matter and space-time
curvature are projected into a lens plane perpendicular to the line
of sight. From one component of the Bianchi identity, we provide a
rigorous, new derivation of a Poisson equation for the projected
matter density where the source term involves second derivatives of
the observed weak gravitational lensing shear. We also show that the
other components of the Bianchi identity reveal no new results.
Numerical integration of the Poisson equation in test cases shows an
accurate mass map can be constructed from the combination of a
ground-based, wide-field image and a Hubble Space Telescope image of
the same system.

\end{abstract}


\maketitle

%________________________________________________________

\section{Introduction} \label{intro:sec}

In the current age of precision cosmology, weak gravitational
lensing is an important tool in understanding cluster mass
morphology.  Because all gravitating mass (baryonic or dark)
influences the path of light rays on the same footing, gravitational
lensing studies are the most direct measure of mass distribution
within galaxies or galaxy clusters.

Current weak gravitational lensing studies relate the projected
matter distribution to the observed gravitational shearing of images
through an integral relation derived in \citet{miralda95} and
elsewhere. Previous work by \citet{seitz} and others introduces a
PDE approach where the gravitational shearing is related first to
the gravitational potential, from which the underlying matter
distribution can be found.

A recent paper by \citet{kk} introduced a new PDE approach to weak
gravitational lensing that directly relates the gravitational shear
to the projected matter distribution.  By examining the weak
gravitational perturbations on a flat background space-time in the
NP spin coefficient formalism for a tetrad of constant null vectors,
\citet{kk} showed that one component of the Bianchi identity
provides a first-order, complex partial differential equation
relating the gravitational shearing of images to the projected
matter density. Using a Green's function, the authors were able to
show that their PDE was equivalent to the integral equation derived
by \citet{miralda95}.

This paper provides a more detailed derivation of the relevant PDE
to first order in gravitational potential and using a non-constant
NP null tetrad. To this order, we show that the other components of
the Bianchi identity (that were not examined in \citet{kk}) yield no
new information related to gravitational lensing.  We show that a
real Poisson equation naturally arises from this complex,
first-order PDE that relates the gravitational shear to the
projected matter density.  The Poisson equation derived here is
related to the Poisson equation of \citet{seitz}, but is derived
from first principles of general relativity.

We numerically integrate the Poisson equation using relaxation
methods for a matter distribution representing a truncated,
isothermal sphere with a core radius. These numerical integrations
are used to study the feasibility of the method for application to
observational scenarios.

%_________________________________________________

\section{NP Formalism} \label{NP:sec}

In this section, we explain the physical arrangement of an observer,
sources and matter distribution that leads to the null tetrad at the
heart of the NP formalism.  We then specify Ricci and Weyl tensor
components and non-zero spin coefficients and list the Bianchi
identities in the NP formalism.

\subsection{Matter distribution and tetrad}

We begin by considering an observer at the origin in a weakly
perturbed, flat Robertson-Walker (RW) space-time.  We assume that
the perturbation is localized in some region far from the observer
and that light sources lie beyond the perturbation, near the line of
sight.  We assume that these sources do not influence the space-time
metric. For simplicity, we take the center of the matter
distribution that perturbs the metric to lie at some distance along
the $-\hat z$ axis. The space-time metric takes the form

\begin{equation}  d \tilde{s}^2 = (1+2 \varphi) \, d\tilde{t}^2
- a(\tilde t)^2 (1-2\varphi) \{ dx^2 + dy^2 + dz^2 \} , \label{m1}
\end{equation}

\noindent where the perturbation $\varphi$ depends on the proper
distance in the background RW metric at the time light rays from the
sources pass the lens on their way to the observer. The metric,
Eq.~\ref{m1} is conformal to the static metric

\begin{equation}  ds^2 = (1+2 \varphi) \, dt^2
- (1-2\varphi)  \{ dx^2 + dy^2 + dz^2 \} , \label{m2} \end{equation}

\noindent and the light rays of the two metrics are identical.

The NP spin coefficient formalism is based on a tetrad of null
vectors,

\begin{equation} \lambda_i^a = \left\{ \ell^a, n^a, m^a, \bar m^a \right\},
\label{tetrad} \end{equation}

\noindent where $m^a$ and $\bar m^a$ are complex spatial vectors
spanning the cross section of a pencil of light rays and $\ell^a$
and $n^a$ are real null vectors. The tetrad vectors satisfy $\ell^a
n_a = 1$ and $m^a \bar m_a = -1$ with all other products equal to
zero. The first null vector, $\ell^a$, is chosen to be tangent to
the past-directed null geodesics.

For our purposes, we will be considering null geodesics that connect
our observer at the origin with sources lying beyond the lens near
the $-\hat z$ axis.  Since we are pursuing weak gravitational
lensing to first order in the gravitational potential, we will
assume that the null geodesics connecting the observer and source
are those of the background metric

\begin{equation} ds^2 = dt^2 -  \{ dx^2 + dy^2 + dz^2 \} .
\label{mb} \end{equation}

\noindent Variations in these null geodesics would be first order
corrections in the gravitational potential. Using the complex
variable

\begin{equation} \zeta = e^{i\phi} \cot{\theta/2} = \frac{x+i\,y}
{\sqrt{x^2+y^2+z^2}-z} \label{zeta} , \end{equation}

\noindent we can write the cartesian coordinate components of null
tetrad vectors as

\begin{eqnarray} \ell^a &=& \frac{1}{\sqrt{2}(1+\zeta\bar\zeta)} \left(
-1-\zeta\bar\zeta, \zeta + \bar\zeta, i(\bar \zeta - \zeta),
-1+\zeta\bar\zeta \right), \label{l} \\
n^a &=& \frac{1}{\sqrt{2}(1+\zeta\bar\zeta)} \left(
-1-\zeta\bar\zeta, -(\zeta+\bar\zeta), i(\zeta-\bar\zeta),
1-\zeta\bar\zeta \right), \label{n} \\
m^a &=& \frac{1}{\sqrt{2}(1+\zeta\bar\zeta)} \left( 0,
1-\bar\zeta^2, -i (\bar\zeta^2 +1), 2\, \bar\zeta \right). \label{m}
\end{eqnarray}

\noindent All of these vectors are null in the background metric,
Eq.~\ref{mb}.  When $\zeta = 0$, the spatial part of $\ell^a$ is
aligned with the $-\hat z$ axis. $n^a$ and $m^a$ are parallel
propagated in the background metric along the $\ell^a$ congruence.
Geometrically, $m^a$ is a complex spatial vector perpendicular to
the spatial part of $\ell^a$ that lies in the cross section of the
light bundle surrounding $\ell^a$.

\subsection{Spin coefficients and curvature tensors}

The Ricci rotation coefficients are defined by

\begin{equation} \gamma_{ijk} = \lambda^b_j \, \lambda^a_k \,
\nabla_a \, \lambda_{i\, b} , \label{ricci_rot} \end{equation}

\noindent and in the NP formalism, these are arranged into twelve
complex NP spin coefficients.  In our examination of the Bianchi
identity, we will assume that $\zeta$ is small so that the light
rays lie close to the $-\hat z$ axis.  The spin coefficients appear
in the Bianchi identity multiplied by a Weyl or Ricci tensor
component which we compute to first order in the perturbation,
$\varphi$.  Hence, we will keep only the zeroth order in $\zeta$
contribution to the spin coefficients.  In the appendix, we show
that the only non-zero spin coefficients to this order are

\begin{equation} \rho = \mu = - \frac{1}{\sqrt{2}r} ,
\label{rho} \end{equation}

\noindent where $r^2 = x^2+y^2+z^2$.

The Ricci and Weyl tensors are decomposed into tetrad components by
contracting the coordinate components with the tetrad vectors. We
assume that the gravitational perturbation and $\zeta$ are small,
and discard terms of the form $\zeta \times \varphi$.  In the
appendix, we find that the non-zero NP Ricci curvature components
are

\begin{equation} \Phi_{00} = \Phi_{22} = 2\Phi_{11} = \frac{1}{2}
\nabla^2 \varphi. \label{ricci} \end{equation}

\noindent The first order Weyl tensor components are

\begin{eqnarray} \Psi_0 = \overline \Psi_4 &=& \frac{1}{2} ( \varphi_{xx} -
\varphi_{yy} - 2i\varphi_{xy} ) \nonumber \\
\Psi_1 = - \overline \Psi_3 &=& \frac{1}{2} (\varphi_{xz} - i
\varphi_{yz}) \nonumber \\
\Psi_2 &=& \frac{1}{2} \left( \varphi_{zz} - \frac{1}{3}\nabla^2
\varphi \right) . \label{weyl} \end{eqnarray}

The four directional derivatives associated with the null tetrad
play an important role in the NP formalism and are given special
names:

\begin{equation} D = \ell^a \partial_a, \quad\quad \Delta = n^a
\partial_a, \quad\quad \delta = m^a\partial_a, \quad\quad \bar\delta
= \bar m^a \partial_a . \label{ders} \end{equation}

\noindent Since the potential is time independent, to zeroth order
in $\zeta$, these derivative operators act as

\begin{equation} D = -\Delta = -\frac{1}{\sqrt{2}} \frac{\partial}{\partial z}
\quad\quad  \delta = \frac{1}{\sqrt{2}}
\left(\frac{\partial}{\partial x} - i \frac{\partial}{\partial y}
\right). \label{ders_z} \end{equation}

\subsection{Bianchi identity}

The Bianchi identity can be expressed in terms of the Ricci and Weyl
tensor components, spin coefficients and the directional derivatives
in eight complex equations \cite{NT}.  A central purpose of this
paper is to examine all of the Bianchi identity equations under the
context of weak gravitational lensing.  Here, we list the Bianchi
identity components that relate the Ricci and Weyl curvature tensor
components, keeping only terms which are first order $\varphi$ and
discarding terms of order $\zeta \varphi$.

\begin{eqnarray}
\bar \delta \Psi_0 - D \Psi_1 - \delta \Phi_{00} & =& -4\rho \Psi_1
\label{A4a} \\
\Delta \Psi_0 - \delta \Psi_1 &=& - \mu \Psi_0 \label{A4b} \\
3(\bar\delta \Psi_1 - D \Psi_2) + 2D\Phi_{11} - \Delta \Phi_{00} & =
& -9 \rho \Psi_2 \label{A4c} \\
3(\Delta \Psi_1 - \delta \Psi_2) - 2 \delta \Phi_{11} & = & -6 \mu
\Psi_1 \label{A4d} \\
3(\bar \delta \Psi_2 - D \Psi_3) + 2 \bar \delta \Phi_{11} & = & -6
\rho \Psi_3 \label{A4e} \\
3(\Delta \Psi_2 - \delta \Psi_3) + D \Phi_{22} -2\Delta\Phi_{11} & =
& -9 \mu \Psi_2 \label{A4f} \\
\bar\delta \Psi_3 - D \Psi_4 &=& -\rho\Psi_4 \label{A4g} \\
\Delta \Psi_3 - \delta \Psi_4 + \bar\delta\Phi_{22} &=& -4\mu\Psi_3
\label{A4h}
\end{eqnarray}

\noindent These equations are trivial identities if one uses the
definitions of the Ricci and Weyl curvature tensor components from
the right hand sides of Eqs.~\ref{ricci} and \ref{weyl}.

Our purpose is to examine these equations as field equations, where
we assume that we know the Weyl tensor components and wish to solve
for the Ricci tensor components.  To this end, we will use the
equivalences of the left hand sides of Eqs.~\ref{ricci} and
\ref{weyl}.

We note that this is the opposite of the program of Newman and
Penrose, who studied gravitational radiation using these equations.
There, vacuum space-time was studied so that the Ricci tensor was
zero, and the Bianchi identities provided field equations to examine
Weyl tensor components at infinity, which represented gravitational
radiation.

%_________________________________________________

\section{Weak gravitational lensing}

In gravitational lensing, the three-dimensional matter density that
leads to the deflection of light is projected into a two-dimensional
surface matter density in the ``lens plane.''  The null geodesics
connecting the observer and a source are replaced by two straight
(asymptotic) rays that form a deflection angle at the lens plane.
This leads to a lens equation mapping image locations in the lens
plane into source positions in the source plane.  The lens equation
map depends on first derivatives (in the lens plane) of the
projected gravitational potential. The Jacobian of this mapping
involves second derivatives of the projected gravitational
potential.  This Jacobian can be used to map the shape of extended
sources in the source plane into extended images in the lens plane,
as is demonstrated in \cite{ehlers} and \cite{fkn2}.

We define $\psi(x,y)$ as the projected gravitational potential,

\begin{equation} \psi(x,y) = \int_o^s \varphi ds, \label{psi}
\end{equation}

\noindent where the integral is taken from the observer position to
the source position along the path connecting the source and
observer. The projected gravitational potential depends only on the
position the ray connecting the observer to the source strikes the
lens plane (whose coordinates are $x$ and $y$).

The goal of weak gravitational lensing is to determine the projected
matter density, $\kappa$, from observed weak gravitational
``shears:'' $\gamma_1$ and $\gamma_2$.  These three quantities arise
in the Jacobian of the lens mapping and are defined as

\begin{eqnarray} \kappa = \frac{1}{2} (\psi_{xx} + \psi_{yy}) & = &
\frac{1}{2} \Delta^2 \psi \label{kappa} \\ \gamma_1 = \frac{1}{2}
(\psi_{xx} - \psi_{yy}) & \& & \gamma_2 = \psi_{xy} . \label{shears}
\end{eqnarray}

\noindent (Note that this matter density $\kappa$ is unrelated to
the spin-coefficient $\kappa$ of Eq.~\ref{spin}.)  Using a Fourier
transform technique, one can show that

\begin{equation} \kappa (\vec{r}) = \int \, d\vec{r}'
\, \frac{[\gamma_1(\vec{r}') \, , \, \gamma_2(\vec{r}') ]}{\pi}
\cdot \frac{ [ \cos 2\eta \, , \, \sin 2\eta ]}{ | \vec{r} -
\vec{r}' |^2} , \label{k} \end{equation}

\noindent where $\eta$ is the angle between the two-dimensional
position vectors $\vec{r}$ and $\vec{r}'$ in the lens plane.  This
integral equation is the basis for observational weak gravitational
lensing.

%_________________________________________________

\section{Projection of Curvature Tensors}

The connection of our NP formalism to gravitational lensing is
accomplished by projecting each of our Bianchi identities and the
relevant curvature components along the paths connecting the
observer and source.  While the Bianchi identity relates curvature
components at each point along the light ray connecting the source
and observer, the projection represents an accumulation similar to
that studied in gravitational lensing.

We begin by projecting the $\Psi_0$ and $\Phi_{00}$ curvature
components.  Since we are interested in weak lensing (where the
deflection is minimal), we project by integrating along the spatial
part of the past light rays given by $\ell^a$ in the tetrad,
Eq.~\ref{l}.  Since we want to work to lowest order in the
potential, this is equivalent to integrating along the $z$ axis from
the observer to the source.  Using a pre-script ``L'' to denote a
quantity projected into the lens plane, we define

\begin{equation} _L\Phi_{00} \equiv  \int_{z_o}^{z_s} \, dz
\, \Phi_{00} = \frac{1}{2} \left( \psi_{xx} + \psi_{yy} \right),
\label{lens_phi00} \end{equation}

\noindent where we use Eq.~\ref{ricci} with Eq.~\ref{psi} and the
assumption that $\varphi_z$ is zero far from the lens.  Likewise,
the Weyl tensor component $\Psi_0$ is projected as

\begin{equation} _L\Psi_0 \equiv  \int_{z_o}^{z_s} \, dz
\, \Psi_0 = \frac{1}{2} \left( \psi_{xx} - \psi_{yy} - 2i \psi_{xy}
\right). \label{lens_psi0} \end{equation}

\noindent We see an immediate connection between the curvature
tensor components and gravitational lensing: $_L\Phi_{00} = \kappa$
and $_L\Psi_0 = \gamma_1 - i \gamma_2$.  In \citet{kk},
$_L\Phi_{00}$ and $_L\Psi_0$ are shown to obey an analogous integral
relation to Eq.~\ref{k}.

The remaining Ricci tensor components project as

\begin{equation} _L\Phi_{22} = 2\,_L\Phi_{11} = \,_L\Phi_{00} =
\kappa. \label{proj:ricci} \end{equation}

\noindent  Likewise, $_L\Psi_4 = \,_L \overline\Psi_0 = \gamma_1 + i
\gamma_2$, but the projections of $\Psi_1$ and $\Psi_3$ are both
zero because both involve a $z$ derivative and we assume $\varphi_x$
and $\varphi_y$ are zero far from the lens. The projection of
$\Psi_2$ is

\begin{equation} _L\Psi_2 \equiv  \int_{z_o}^{z_s} \, dz
\, \frac{1}{2} \left( \varphi_{zz} - \frac{1}{3}\nabla^2 \varphi
\right) = -\frac{1}{6} \left( \psi_{xx} + \psi_{yy} \right) =
-\frac{1}{3} \kappa .\label{proj:psi2} \end{equation}

\noindent That the $_L\Psi_2$ Weyl tensor component is equal to the
mass density is consistent with the role $\Psi_2$ plays in the Bondi
definition of the mass of an asymptotically flat space-time.

%_________________________________________________

\section{Projection of Bianchi identities}

Each Bianchi identity, Eqs.~\ref{A4a}-\ref{A4h}, can be projected in
a similar manner.  While each identity holds under the projection
when using the right hand sides of Eqs.~\ref{ricci} and \ref{weyl},
our interest is in forming field equations for the components of the
Ricci tensor (or matter density) given the Weyl tensor (or
gravitational shear) as a source under the conditions of
gravitational lensing. This justifies throwing away boundary value
terms or terms that are ``small'' under the projection relative to
other quantities in the equation.

Under the projection, $\delta$ and $\bar\delta$ directional
derivatives lie in the lens plane (to zeroth order in $\zeta$) and
the integration can be passed through the derivative. This implies,
for example, that the projection of $\delta \Psi_1$ will be zero
since the projection of $\Psi_1$ is zero.  When the projection is
applied to quantities on which the $D$ and $\Delta$ derivative
operators act, one has

\begin{equation} \int_{z_o}^{z_s} \, dz \, D \, \Phi_{11} = -
\frac{1}{\sqrt{2}} \int_{z_o}^{z_s} \, dz \frac{\partial}{\partial
z} \, \Phi_{11} \approx 0, \end{equation}

\noindent where we will assume that all the curvature tensors are
approximately zero far from the lens.

For four of the Bianchi identities, Eqs.~\ref{A4b}, \ref{A4c},
\ref{A4f}, and \ref{A4g}, the entire left hand side of the Bianchi
identity ``vanishes'' under projection.  This means that these
equations have no content as field equations for projected Ricci or
Weyl tensor components.  The equality of the identity under the
projection holds exactly because the small boundary value terms that
we choose to neglect cancel the small projected term on the right
hand side of the equation.

Using Eqs.~\ref{ricci}, \ref{weyl} and \ref{ders_z}, one can see
that the four remaining Bianchi identities represent two independent
equations; Eq.~\ref{A4d} is equivalent to Eq.~\ref{A4e}, and
Eq.~\ref{A4a} is equivalent to Eq.~\ref{A4h}. Under projection,
these equations become

\begin{eqnarray} \delta \left( 3\,_L\Psi_2 + 2\,_L\Phi_{11} \right)
& = & 6  \int_{z_o}^{z_s} \, dz \, \rho\Psi_1 , \label{A4d1} \\
\delta \,_L\Phi_{00} - \bar\delta \,_L\Psi_0 & = & 4
\int_{z_o}^{z_s} \, dz \, \rho\Psi_1 . \label{A4a1}
\end{eqnarray}

Recall that our physical setup was to have a lens at $z=z_l$ and the
observer at the origin.  At the observer, $\rho$ diverges because
the past light-cone which generates the null tetrad has its apex at
the observer. For physically reasonable gravitational potentials of
the form $\varphi(|\vec{r}-\vec{r_l}|)$ far from the lens, the
combination $\rho \Psi_1$ is non-zero but finite and small at the
observer's location. In addition, because the physical gravitational
potential will only be substantially different from zero near the
lens, the combination $\rho\Psi_1$ is a nearly odd function in
$z-z_l$ due to the $z$ derivative in the definition of $\Psi_1$.
Thus, the projection of $\rho \Psi_1$ is very small compared with
individual terms on the left hand side of Eqs.~\ref{A4d1} and
\ref{A4a1} for physically reasonable models.

Neglecting these small terms, we see that Eq.~\ref{A4d1} ceases to
be a differential equation:

\begin{equation} \delta \left( 3\,_L\Psi_2 + 2\,_L\Phi_{11} \right)
= 0 \Rightarrow 3\,_L\Psi_2 + 2\,_L\Phi_{11} = c \end{equation}

\noindent for some constant $c$.  Using the definitions of the
projection, Eqs.~\ref{proj:ricci} and \ref{proj:psi2}, we see that
the constant is zero, and this equation has no implications for
gravitational lensing.

The remaining Bianchi identity becomes

\begin{equation}
\delta \, _L\Phi_{00}  =   \bar\delta \, _L\Psi_0, \label{BI1}
\end{equation}

\noindent which is a field equation for the projected mass density,
$_L\Phi_{00}$ or $\kappa$, with the observed gravitational shear,
$_L\Psi_0 = \gamma_1-i\gamma_2$, as a source.  Eq.~\ref{BI1} was the
field equation reported in \citet{kk}, so that the different choice
of tetrad between the two papers was not important. In \citet{kk},
it is shown that Eq.~\ref{BI1} is equivalent to the integral
equation, Eq.~\ref{k}, by use of a Green's function on the sphere.

By applying the $\bar\delta$ operator to both sides of
Eq.~\ref{BI1}, a Poisson equation is obtained:

\begin{equation} \bar\delta \delta \,_L\Phi_{00} = \frac{1}{2}
\left(\frac{\partial^2}{\partial x^2} + \frac{\partial^2} {\partial
y^2} \right) \, _L\Phi_{00} =  \bar\delta^2\,_L\Psi_0.
\label{poisson1}
\end{equation}

\noindent  This partial different equation provides a
straight-forward way to determine gravitational lensing mass maps
from observed weak lensing shear.  Equation~\ref{poisson1} is fully
accurate and consistent to first order in the gravitational
potential.  The source term, $\bar\delta^2 \, \Psi_0$ is manifestly
real when written in terms of the gravitational lensing observables
$\gamma_1$ and $\gamma_2$. Using $_L\Psi_0 = \gamma_1 - i \gamma_2$,
we have

\begin{equation} \bar\delta^2 \, _L\Psi_0 = \frac{1}{2}
\left( \gamma_{1xx} - \gamma_{1yy} +
2\gamma_{2xy} \right) - \frac{i}{2} \left( \gamma_{2xx} -
\gamma_{2yy} - 2\gamma_{1xy} \right). \label{poisson2}
\end{equation}

\noindent  Using the definitions of $\gamma_1$ and $\gamma_2$
(Eq.~\ref{shears}), one can show that the imaginary part of
Eq.~\ref{poisson2} will always be zero.  Hence, the component of the
projected Bianchi identity, Eq.~\ref{BI1}, informs us that projected
mass density is related to the weak lensing shears by

\begin{equation} \Delta^2 \,_L\Phi_{00} = \Delta^2 \kappa = \kappa_{xx} + \kappa_{yy} =
2 \bar\delta^2 \, _L\Psi_0 = \gamma_{1xx} - \gamma_{1yy} +
2\gamma_{2xy}. \label{poisson3}
\end{equation}

\noindent  Hereafter, we will refer to the projected Ricci and Weyl
tensor components $_L\Phi_{00}$ and $_L\Psi_0$ as the projected
matter density and gravitational shears, $\kappa$ and
$\gamma_1-i\gamma_2$.

%______________________________________________________

\section{Numerical Methods}

We examine the feasibility of integrating Eq.~\ref{poisson3} by
introducing a sample matter distribution and computing $\kappa$
directly from Eq.~\ref{proj:ricci} and indirectly from numerical
integration. Our goal is to evaluate, under the simplest
observational scenarios possible, whether numerical integration of
Eq.~\ref{poisson3} is a promising approach to determining accurate
mass maps.  Keeping with the simple approach, we choose to use
relaxation methods for our numerical integration based on
\citet{nrc}.

In relaxation methods, one attempts to set a grid spacing that is
appropriate to the rate of change of the functions being integrated.
However, we choose to constrain the grid spacing based on
approximate number counts of objects in the background of a lens
that are resolved by the telescope in question. Observational weak
gravitational lensing is an inherently statistical study since the
natural shape of unlensed background objects is not known and only
assumed to be {\emph{on average}} circular.  For the purposes of
this paper, we assume that there are not intrinsic background galaxy
alignments.

Far from the center of a matter distribution, the gravitational
lensing shear signal is weak, and a relatively large number of
objects is required to form a good estimate of the underlying shear.
Closer to the center of the distribution, a smaller number of
objects is needed because the shear is substantially larger.
Generally, 30-40 objects far from the center of the matter
distribution and 10 objects in the central region are sufficient for
a reasonable statistical weak lensing analysis.

For the purposes of our numerical tests, we consider that one can
measure average values of $\gamma_1$ and $\gamma_2$ in each box and
assign these to grid values. While we could compute analytic
derivatives of $\gamma_1$ and $\gamma_2$ to use on the right hand
side of Eq.~\ref{poisson3}, we will use numerical derivatives based
on the grid values.  Hence, as the grid spacing grows, precision
will be lost in the source term of our Poisson equation.  This is
appropriate because we are attempting to model actual practice where
the data must be binned to overcome the statistical nature of the
measurements.

For our model, we consider a truncated isothermal sphere matter
distribution with a core radius, whose three dimensional matter
density is given by

\begin{equation} \rho = \frac{\sigma_v^2}{2\pi\,G}
\frac{1}{r_c^2+r^2} \frac{r_t^2}{r_t^2+r^2}, \label{3drho}
\end{equation}

\noindent where $\sigma_v$ is the velocity dispersion and $r_c$ and
$r_t$ are the core and tidal radii of the distribution.  The model
is inspired by the truncated NFW model proposed by \citet{baltz},
but yields simpler analytic expressions for the projected matter
density and gravitational shear expressions than either NFW or
truncated NFW models. For $r \ll r_t$, the model represents an
isothermal sphere with a core radius. The total mass of the model
(extended to infinity) is finite and is given by

\begin{equation} M_{tot} = \frac{\pi r_t^2 \sigma_v^2}{G\,(r_c+r_t)},
\end{equation}

\noindent and the three-dimensional gravitational potential
approaches a point mass potential for large $r$.

The projected mass density is

\begin{equation} \Sigma = \int_{-\infty}^{\infty} \, dz \rho =
\frac{\sigma_v^2}{2\,G}\frac{r_t^2}{r_t^2-r_c^2} \left(
\frac{1}{\sqrt{s^2+r_c^2}} - \frac{1}{\sqrt{s^2+r_t^2}} \right),
\label{sigma:tis} \end{equation}

\noindent for the radial parameter $s = \sqrt{x^2+y^2}$ in the lens
plane.  The projected gravitational potential, $\psi$ found by
integrating the two-dimensional Poisson equation and using the
constants of integration to make the potential finite at the origin
is

\begin{equation} \psi = -\frac{2\pi\sigma_v^2\,r_t^2}{r_t^2-r_c^2}
\left[ \sqrt{s^2+r_t^2} - \sqrt{s^2+r_c^2} + r_c \ln \left( \frac{
r_c + \sqrt{s^2+r_c^2}}{r_c} \right) - r_t \ln \left( \frac{ r_t +
\sqrt{s^2+r_t^2}}{r_t} \right) \right]. \label{psi:tis}
\end{equation}

The Ricci tensor or $\kappa$, can be explicitly calculated from
Eqs.~\ref{proj:ricci} as

\begin{equation} \kappa = -\frac{\pi\sigma_v^2\, r_t^2}{r_t^2-r_c^2} \left(
\frac{1}{\sqrt{s^2+r_t^2}} - \frac{1}{\sqrt{s^2+r_c^2}} \right)
\label{ricci:tis} .\end{equation}

\noindent The gravitational shears for this matter distribution are

\begin{eqnarray} \gamma_1 &=& - \left( \frac{\sigma_v^2 \pi r_t^2}
{r_t^2 - r_c^2} \right) (y^2-x^2) {\mathcal{Q}}, \nonumber \\
\gamma_2 &=& \left(\frac{\sigma_v^2 \pi r_t^2} {r_t^2 -
r_c^2}\right) (2xy) {\cal{Q}}, \label{shears:tis}
\end{eqnarray}

\noindent for

\begin{equation} {\cal{Q}} = \frac{1}{\sqrt{s^2+r_t^2}(r_t
+ \sqrt{s^2+r_t^2})^2} - \frac{1}{\sqrt{s^2+r_c^2}(r_c +
\sqrt{s^2+r_c^2})^2}, \label{Q}
\end{equation}

\noindent and cartesian coordinates $\{x, y\}$ in the lens plane. We
will compute $\gamma_1$ and $\gamma_2$ from Eq.~\ref{shears:tis} and
assign these values to each grid point.  We emphasize that we will
take numerical derivatives of these values to simulate the case of
observational data. We note that Poisson's equation is independent
of $\sigma_v$, so that the relative errors we cite in this paper are
invariant under scaling $\sigma_v$, which controls the mass of the
cluster.

For our numerical modeling, we choose to set the core radius to
$250$~kpc and the tidal radius to $1.5$~Mpc.   The total mass of the
cluster for a velocity dispersion of $1500$~km~s$^{-1}$ is equal to
$2.1\times10^{15}$ solar masses.  At a radius of $0.5$~Mpc from the
center, the ratio of the densities $\rho(r=0.5$Mpc$)/\rho(r=0)$ is
only $0.18$, but $6.0$~Mpc from the center, this ratio has fallen to
$1.0\times10^{-4}$.

%__________________________________________________________

\section{Numerical Tests}

In this section, we are interested in testing the integration of the
Poisson equation, Eq.~\ref{poisson3}, under conditions
representative of typical weak lensing measurements.  We consider
ground based observations by wide field telescopes, space based
observations made by HST, and a hybrid, two-grid combination where
one has both a quality ground and space based image.

As our primary comparison, we look at the scaled error in the
projected mass density in each box, found by subtracting the value
of $\kappa$ found by numerical integration of Eq.~\ref{poisson3} to
the exact expression in Eq.~\ref{ricci:tis} and dividing by the
correct value of $\kappa$ at the center. (The more common relative
error is less meaningful, as it approaches one at the edges of the
grid because we set the edge values to zero as the boundary
condition, even though this is in fact a small error.) We also
compare the total integrated masses, which we estimate as the sum of
the mass densities in each box times the box area.

As an example of a standard ground-based, wide-field telescope
usage, we consider the number counts and field of view (fov) for the
DLS project reported in \citet{wittman}.  Here, 4-m class telescopes
with 35 arc sec fov were used to observe five well separated fields
for weak lensing signals.  Filtering on well observed objects using
the standard software for source extraction (SExtractor
\cite{bertin}) and shape measurement (ellipto \cite{bernstein}),
yielded 20 objects per square minute. Assuming that a lens is at a
redshift of $0.45$ with minimal dithering of the stacked images,
this translates into an observed lens plane that is approximately
$12$~Mpc across in the standard cosmology.

Hence, while one would like a smaller grid spacing for resolution,
dividing the field of view into $25\times25$ boxes that are
$0.48$~Mpc across yields an appropriate average of $40$ objects per
box. Adding one out of field of view set of data points to which one
can assign a boundary value of zero yields a $27\times27$ grid for
relaxation of Eq.~\ref{poisson3}.  We note that when computing
numerical derivatives along the edges of the interior $25\times25$
grid, it is appropriate to use only the interior (measured data)
points and not the added exterior edge.

\begin{figure}
\begin{center}
\scalebox{0.8}{\includegraphics{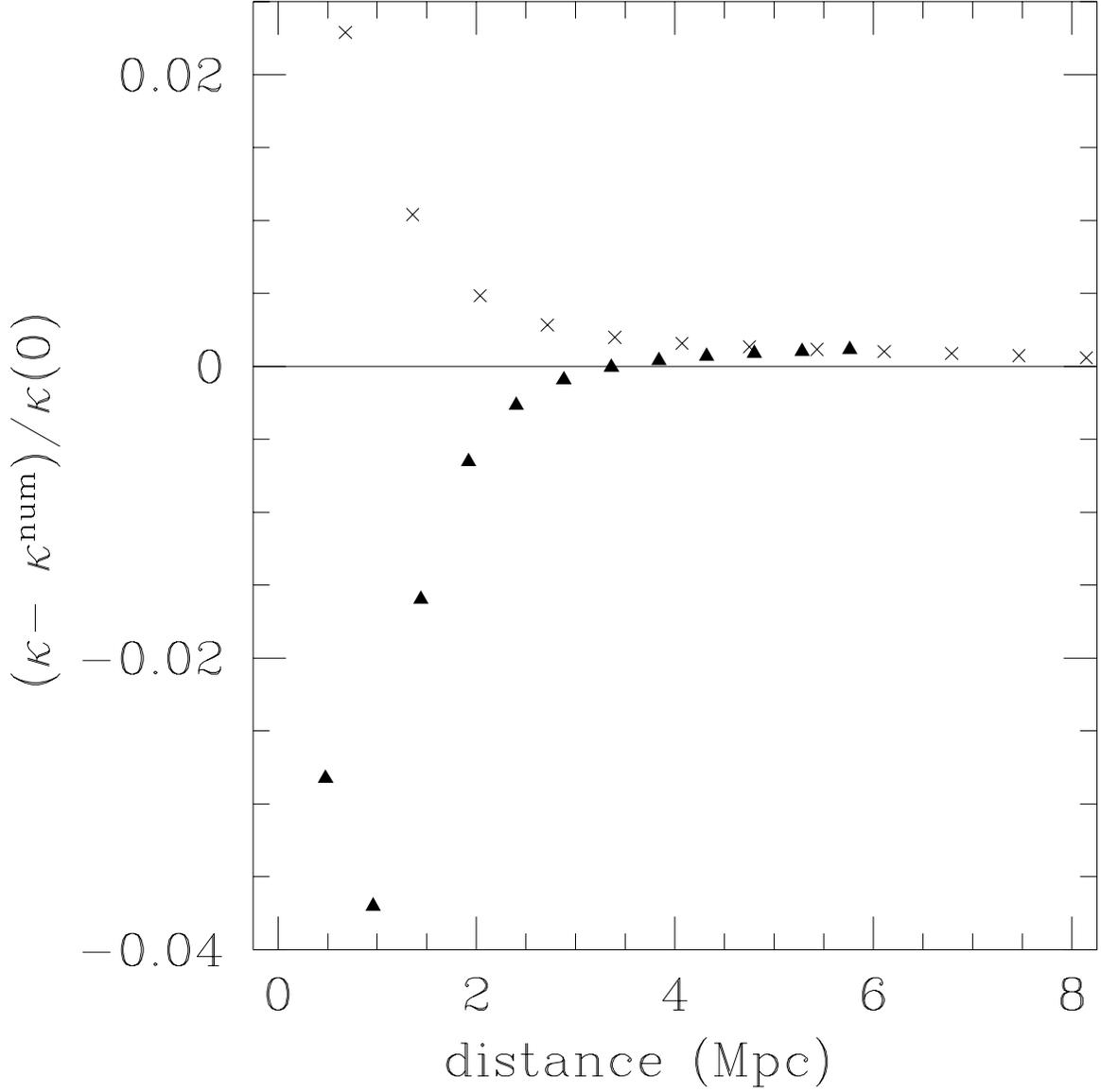}} \caption{\label{wide:fig}
The error in the projected mass density scaled by the value of
$\kappa$ at the origin for a wide field image with grid separation
of $0.48$~Mpc. The triangles represent the grid points along the $x$
axis passing through the origin, while the crosses represent the
grid points along the diagonal.}
\end{center}\end{figure}

Figure~\ref{wide:fig} shows the scaled error in the projected mass
density computed between numerical integration of Eq.~\ref{poisson3}
and the exact value in Eq.~\ref{ricci:tis} when the boundary of the
$27\times27$ grid is set to zero.  The error in the central bin (not
shown in Figure~\ref{wide:fig}) is $+31\%$, but the errors in other
bins are all small. Along the $x$ axis passing through the center of
the distribution, the numerical integration slightly overestimates
the mass density for values near the center, whereas along the
diagonal, the numerical integration always slightly underestimates
the correct value. When the boundary values are set to the correct
values, which one would never know in practice, there is no
significant change in the error estimates, implying that assigning
zero to the boundary is not a significant source of error. The total
integrated mass, estimated as the sum of the mass density times the
box area, from the numerical integration underestimates the total
integrated mass by only $8.8\%$. This is in part because while the
central mass density value is significantly underestimated, the
numerical integration overestimates other boxes.

By setting the grid spacing to an arbitrarily small value, one can
achieve good resolution.  Using a $101 \times 101$ boxes for this
field of view, corresponds to a grid spacing of $~0.118$~Mpc.  For
this spacing, the maximum relative error is in the central bin and
is only $3.3 \%$, and the total integrated mass is correct to
$7.8\%$. Plots of the scaled error for this number of boxes has the
same general features as Fig.~\ref{wide:fig}.  Setting the added
boundary values to the (in principle unknown) correct values results
in extremely accurate numerical integration at this number of boxes,
as one would expect.

Therefore, we see that the overall limitation is the requirement of
large boxes.  For the scenario we have described, using $101 \times
101$ boxes results in only $2.4$ objects per box. We conclude that
for ground-based, wide-field telescopes under normal operating
conditions, numerical integration of Eq.~\ref{poisson3} by
relaxation with large boxes will be good enough to detect weak
lensing and get a sense of the overall mass, but will be
insufficient for detailed mass maps.

We also see that the mass sheet degeneracy is not a major issue for
wide field images.  Setting the edges to zero (as opposed to the
unknown correct values) does not significantly effect the results.
This is because the ratio of mass density at the edge of a wide
field compared to the center is small.

For a space-based optical telescope such as the HST, a significantly
higher number of background objects can be resolved, but the field
of view is substantially smaller.  Based on the HUDF number counts
presented in \citet{beckwith}, one can expect 150 objects per square
arc minute for the same cuts as are used for the DLS sample.  The
HST field of view is approximately $200$ arc sec across, so that for
a lens at a redshift of $0.45$, the lens plane is only $1.14$~Mpc by
$1.14$~Mpc.  However, HST can resolve approximately $1100$
background objects in this field of view. Dividing the HST image
into $11\times11 = 121$ boxes results in a grid spacing of
$~0.104$~Mpc with approximately $13$ objects per box.

\begin{figure}
\begin{center}
\scalebox{0.8}{\includegraphics{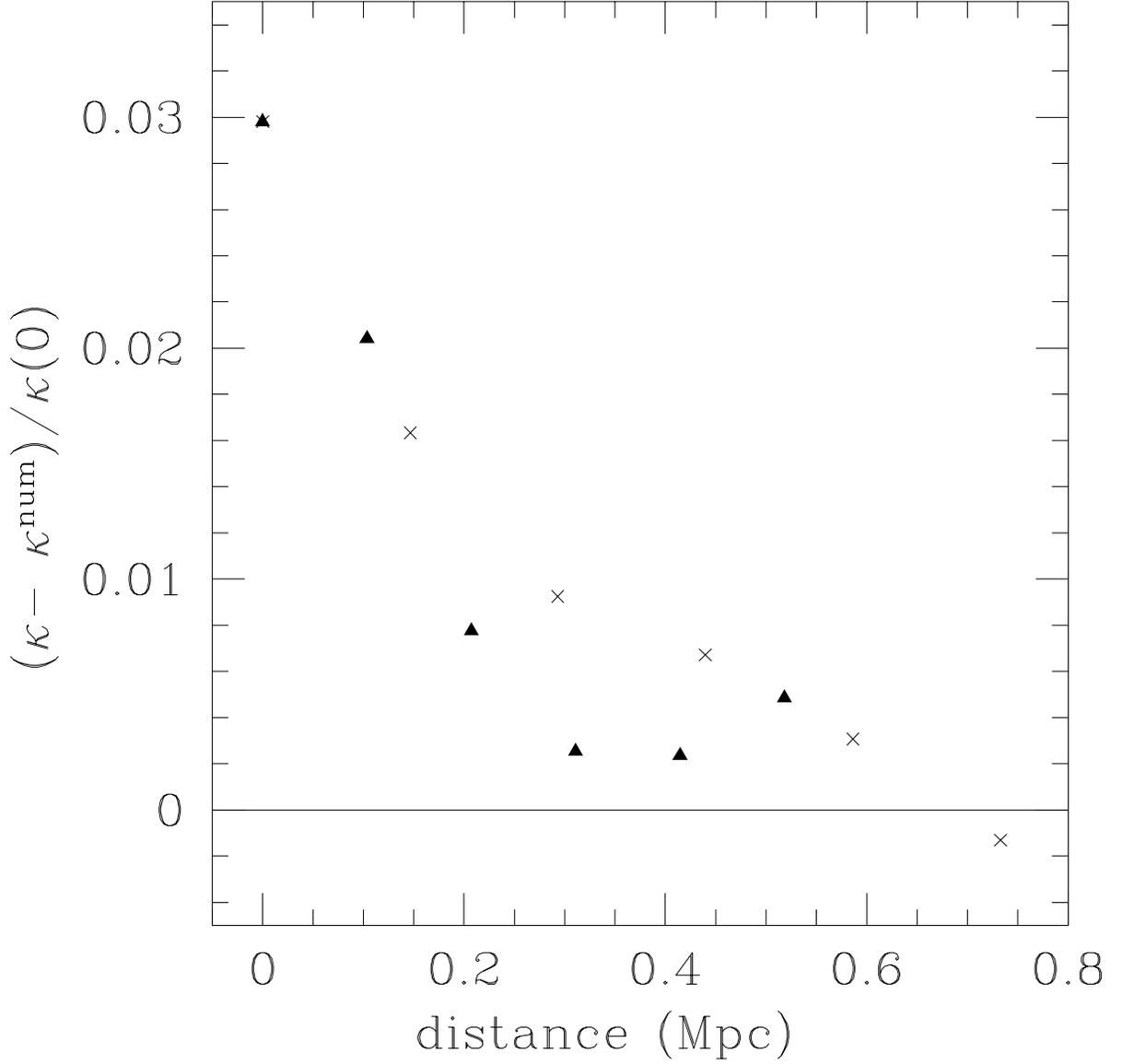}} \caption{\label{hst:fig}
The error in the projected mass density scaled by the value of
$\kappa$ at the origin for an HST image with grid separation of
$0.104$~Mpc. The boundary of the grid has been set to the (unknown)
correct values. The triangles represent the grid points along the
$x$ axis passing through the origin, while the crosses represent the
grid points along the diagonal.}
\end{center}\end{figure}

Because the mass density $0.5$~Mpc from the center of a cluster is
not close to zero, poor results are achieved when relaxing with zero
as a boundary value.  To see what one could achieve, we set the
boundary values to the correct values and use relaxation to find the
projected mass density. Figure \ref{hst:fig} shows that the scaled
error indicates that good modeling is possible if one knows the
boundary values, with a maximum relative error of $3.0\%$.  The
scaled error curves have similar, but more gentle slopes to those of
the wide-field telescopes.  The total relative error in the mass is
$1.3\%$.

This modeling is unphysical because one can never know the true mass
density to assign at the boundary of the HST image.  However, if one
had both a ground-based, wide field image and an HST image of the
same cluster, then one could use the wide-field image to give
interpolated values for the boundary of the HST image.

To demonstrate the effectiveness of this approach, we assume that we
have a wide field, ground based image and an HST image of the same
system with the parameters above.  We assume that the centers of
these two images are the same and that there is no rotation between
the images.  After relaxing the wide field image using $25\times25$
boxes as before, we use the numerically obtained values of $\kappa$
from the large grid to find values for the boundary of the HST image
by interpolation.  We then relax the small grid with these
interpolated boundary values.

\begin{figure}
\begin{center}\scalebox{0.8}{
\includegraphics{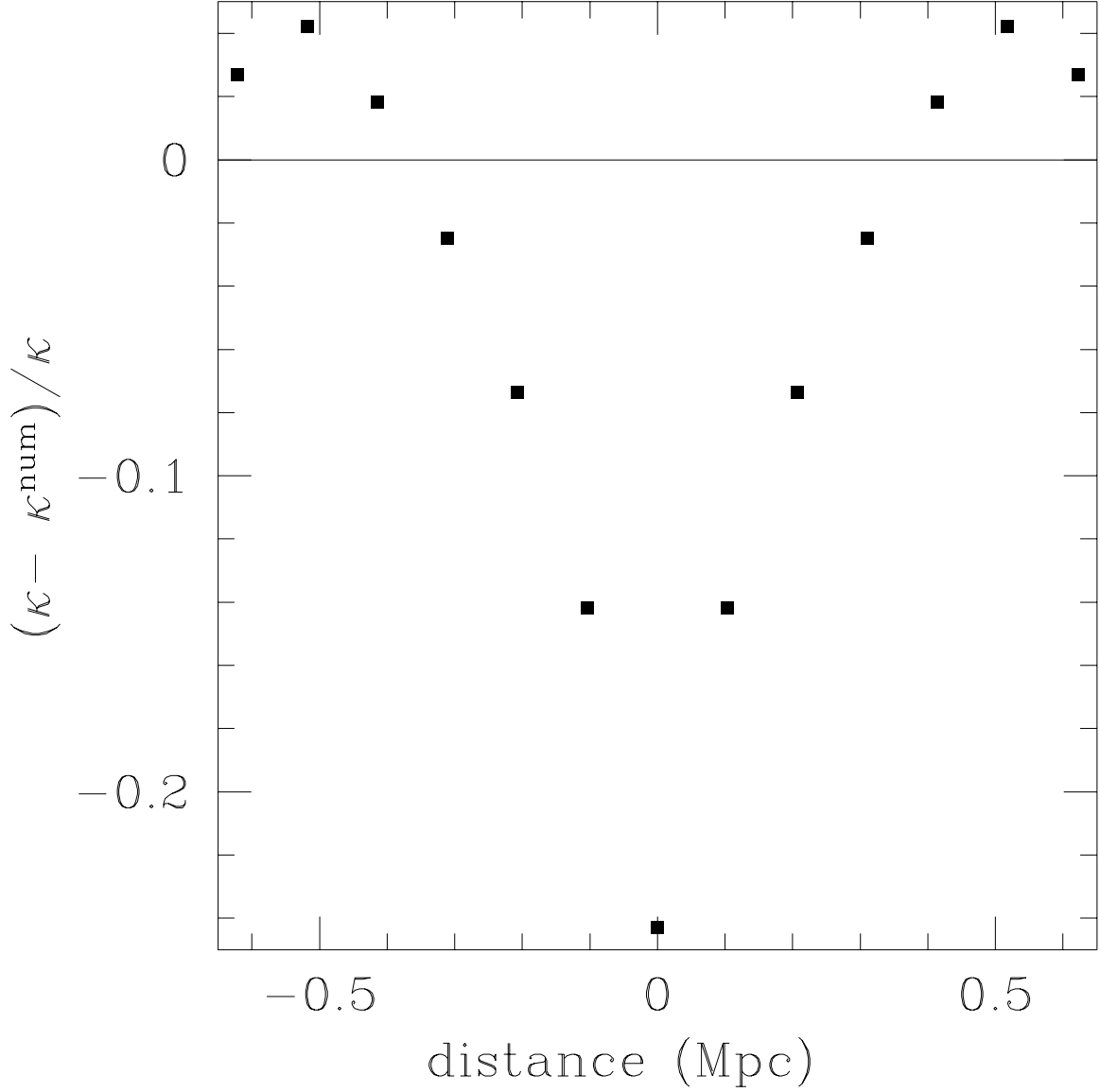}}
\caption{\label{border:fig} The relative error in the projected mass
density along the boundary of an HST image between the true values
and values interpolated from a wide field image with a grid spacing
of $0.48$~Mpc.} \end{center} \end{figure}

The relative error in the $\kappa$ boundary values is shown in
Fig.~\ref{border:fig} along one line of the boundary of the small
field of view grid.  The points at large $x$ correspond to the
corners of the grid, and by design, the errors are symmetrical along
each of the four edges of the small grid.  We see that the
approximate boundary values have a moderate relative error, where
the relaxed values from the large grid slightly underestimate the
true values near the corners of the grid by about $5\%$ and
substantially overestimate the values near the center of the grid by
about $25\%$.

Even though the relative errors in the small grid boundary values
are moderate, the actual errors are small compared to assigning zero
on the boundary of the small grid.  This results in successful
relaxation of the small grid in reproducing the mass density. The
error in the central bin is less than $1\%$ and the total relative
error in the predicted mass on the small grid is $-3.1\%$. (The
relaxed total mass is more than the actual value.)

\begin{figure} \begin{center} \scalebox{0.8}{
\includegraphics{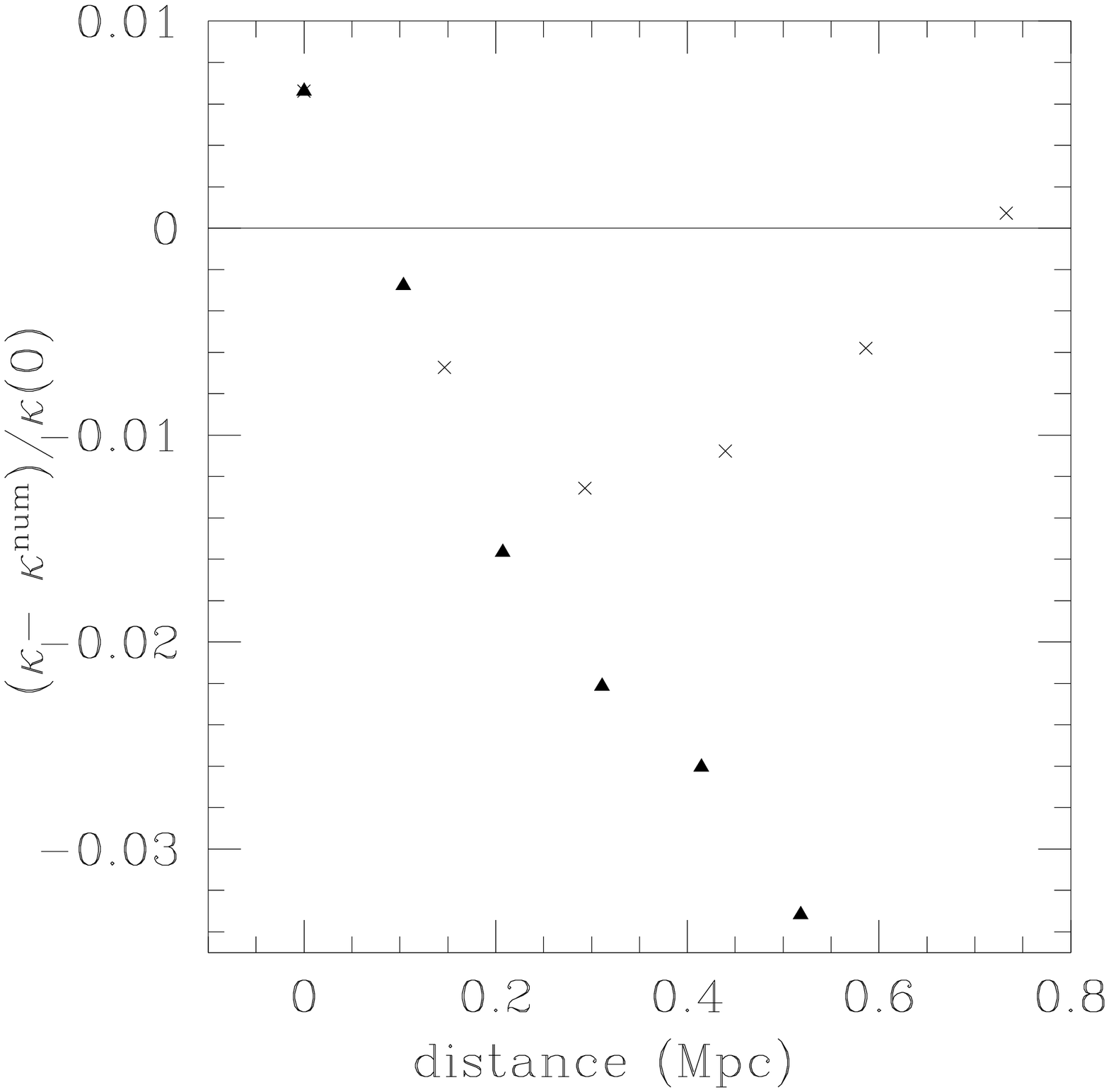}}
\caption{\label{2grid:fig} The error in the projected mass density
scaled by the value of $\kappa$ at the origin for an HST image with
grid separation of $0.48$~Mpc where the boundary values are
determined by interpolation from a wide field image.  The triangles
represent the grid points along the $x$ axis passing through the
origin, while the crosses represent the grid points along the
diagonal. }
\end{center}\end{figure}

Figure~\ref{2grid:fig} shows the scaled errors along the $x$ axis
and diagonal as before.  These scaled errors clearly track the error
in the boundary conditions found by interpolation, as one would
expect. Nevertheless, the errors are all acceptably small so that
one can use the ground based image to provide boundary values and
the space based image to provide accurate mapping of the mass
density.

%________________________________________________________

\section{Discussion}

This paper includes a rigorous analysis of the Bianchi identity
under the conditions of gravitational lensing leading to the
derivation of a Poisson equation for the projected mass density in
terms of derivatives of the observed lensing shear. Using the
simplest possible numerical integration scheme, we show that it may
be possible to use the Poisson equation to determine accurate mass
maps when a ground-based, wide-field image and a high-resolution,
space-based image of the same system are used in a two grid method.

Compared with an initial investigation of the Bianchi identity
\cite{kk}, this paper uses a more physical tetrad that makes
manifest the connection to the light cone of the observer.  In this
way, the paper is more in the spirit of the non-perturbative
approach initiated by \citet{fn} and independently by
\citet{perlick}.  The resulting differential relation,
Eq.~\ref{BI1}, found in both papers is derived here in a more
rigorous fashion where all the relevant physical approximations are
made.

One might have hoped that the other components of the Bianchi
identity, not examined previously in this context, would have
yielded new differential relations between observables.  In
particular, one might have suspected that a Bianchi identity
component might have contained information regarding image
magnification, which can also be tied to the mass density.  We show
that this is not the case, and that no new physical information is
contained in the remaining seven Bianchi identity components.

A full non-perturbative treatment of image distortion is given in
\citet{fkn2}.  It is well known that the optical scalars, $\rho$ and
$\sigma$, which control the divergence and shearing of the pencil of
rays connecting an extended source to the observer through the
optical scalar equation, are related to the components of the Ricci
and Weyl tensors through the Sach's equation.  \citet{fkn2} show in
detail how this leads to image distortion (both shearing and
magnification).

The advance of this paper is to explicitly derive a relation between
the mass density (represented by the Ricci tensor component
$_L\Phi_{00}$) and the gravitational shears (represented by the Weyl
tensor component $_L\Psi_0$).  This connection is the meaningful
observational relationship, but the fact that it comes directly from
the Bianchi identity, rather than a manipulation of the optical
scalar equations or the Sach's equation, would have to be viewed as
a happy accident.

The Poisson equation derived here is related to the Poisson equation
reported in \citet{seitz}, which is derived kinematically by
considering the relationships between $\kappa$, $\gamma_1$, and
$\gamma_2$ and the underlying projected gravitational potential.
When the terms that are not linear in the gravitational potential
are removed from the Poisson equation in \citet{seitz}, one obtains
the Poisson equation reported here.  Removing the non-linear terms
is consistent with the linearized metric used here.

We examined reasonable observational scenarios in this paper and
found that for a wide-field image alone, it would at least be
difficult to use Eq.~\ref{poisson3} to determine accurate mass maps.
Here, we studied only relaxation methods, which require that the
data be binned to achieve an acceptable average shear.  It is
possible that the more complicated to implement method of maximum
likelihood, which can use individual data points, would result in an
accurate mass map.

We were able to show that the combination of a ground-based,
wide-field image and an HST image could be used to determine an
accurate mass map in the central HST region.  The fact that good
results were achieved using relaxation methods indicates that the
numerical integration of Eq.~\ref{poisson3} is a promising road for
weak lensing mass maps.  When using a wide-field image, we examined
the mass sheet degeneracy and found that knowing the correct
boundary values did not significantly improve the accuracy of the
mass maps.

Our paper did not consider several important sources of error.
First, we did not add noise to the shear measurements, which would
more accurately reflect the observational situation.  The presence
of small noise would make the errors in the two grid method larger,
but we believe that our large box size (large enough to accommodate
roughly 20 objects per box), would minimize this error.  Second, we
did not model missing data in our analysis.  Bright nearby stars
lead to saturation in regions on the ccd image, obscuring sometimes
large regions and generally reducing the ability to extract
background objects nearby.  This would both add to the noise in the
average boxed shear values and typically cause a handful of boxes to
have no measured data.  One could model these missing data points
using surrounding data points if this was the case.  Overall, we
expect that these two sources of error would lead to a decrease in
the effectiveness of the numerical integration, but would not cause
the numerical integration to fail.

Two minor issues not examined here related to the ground and HST
images are not expected influence the results of our study.  First,
the two images are unlikely to be exactly centered.  Second, there
is typically a rotation between the two images.  Using common stars
or bright objects, one can easily orient the two coordinate systems.
It is then simply a matter of interpolating from the large grid onto
the small grid boundary.  Since ground based images are so much
larger than HST images, the center offset would have a minimal
impact. A rotation would cause the relative error of
Fig.~\ref{2grid:fig} to vary in position, but the overall
effectiveness should be maintained.

%_______________________________________________________

\section{Conclusion}

This paper gives a rigorous derivation of a Poisson equation,
Eq.~\ref{poisson3}, that directly relates the mass density to
derivatives of the weak lensing shears.  We show that no new
information is obtained by examining the remaining components of the
Bianchi identity, so that the full theory of weak gravitational
lensing and image distortion is most completely described by the
optical scalar equations discussed in \citet{fkn2}.

Using the simplest possible integration scheme, we show that the
Poisson equation can be used to detect weak lensing signals, but
fails to give accurate mass maps for wide-field, ground based
images.  However, simple relaxation methods were shown to be
successful when a ground and HST image were both present.  This
method of determining the matter distribution could prove very
useful in developing highly accurate mass maps.

Further examination of the Poisson equation is warranted.
Specifically, it would be interesting to know whether maximum
likelihood methods could yield accurate mass maps for ground based
images alone.  Also, the application of the two grid method to mock
background images that include appropriate levels of noise would
solidify the usefulness of the method.

%__________________________________________________

\begin{acknowledgments}
BC thanks the Bridgewater State College Adrian Tinsley Program for
Undergraduate Research for a Summer Grant that enabled his
participation in this project.  We acknowledge Ren Li, whose initial
examination of the Bianchi identity components during a senior
independent study directed some of our results, and Dr. Ian
Dell'Antonio of Brown University for helpful discussions related to
number counts of the DLS survey.
\end{acknowledgments}

\appendix*

\section{NP Formula}

For completeness, in this appendix we briefly outline the
calculation of the NP tetrad components of the Ricci and Weyl
tensors and the spin coefficients.  For the perturbed metric given
by Eq.~\ref{m2}, the non-zero, first order Ricci and Weyl tensor
components are given by

\begin{equation} R_{00} = - \nabla^2 \varphi \quad\quad R_{ii} =
-\nabla^2 \varphi, \label{ricci:c} \end{equation}

\noindent with $\nabla^2 = \partial_x^2 + \partial_y^2 +
\partial_z^2$, and

\begin{eqnarray} C_{0i0i} = \frac{1}{3} \left( -3 \varphi_{ii} +
\nabla^2 \varphi \right) & \quad\quad & C_{0i0j} = -\varphi_{ij}
\quad i \ne j \nonumber \\ C_{ijij} = \frac{1}{3} \left( 3
\varphi_{kk} - \nabla^2 \varphi \right) \quad i\ne j \ne k &
\quad\quad & C_{ijik} = -\varphi_{jk} \quad i \ne j \ne k,
\label{weyl:c} \end{eqnarray}

\noindent where we use $0$ for time and $i,j,k$ for spatial
components, ignoring the usual index conventions.

The NP tetrad components of the Weyl tensor are defined by

\[\Psi_0 = -C_{abcd} \ell^a m^b \ell^c m^d ,
 \quad\quad  \Psi_1 = -C_{abcd} \ell^a n^b \ell^c m^d , \]
\[ \Psi_2  =  -\frac{1}{2} \left( C_{abcd} \ell^a n^b \ell^c n^d
- C_{abcd} \ell^a n^b m^c \bar m^d \right)  , \]
\begin{equation} \Psi_3 = C_{abcd} \ell^a n^b n^c \bar m^d ,  \quad\quad
\Psi_4 = -C_{abcd} n^a \bar m^b n^c \bar m^d, \label{weyl_np}
\end{equation}

\noindent and the tetrad components of the Ricci tensor are

\begin{eqnarray} \Phi_{00} = -\frac{1}{2} R_{ab} \ell^a \ell^b ,
\quad\quad & \Phi_{10} = -\frac{1}{2} R_{ab} \ell^a \bar m^b , &
\quad\quad \Phi_{20} = -\frac{1}{2} R_{ab} \bar m^a \bar m^b ,
\nonumber \\
\Phi_{01} = -\frac{1}{2} R_{ab} \ell^a m^b , \quad\quad & \Phi_{11}
= -\frac{1}{4} \left( R_{ab} \ell^a n^b + R_{ab} m^a \bar m^b
\right) , & \quad\quad \Phi_{21} = -\frac{1}{2} R_{ab} n^a \bar m^b
,
\nonumber \\
\Phi_{02} = -\frac{1}{2} R_{ab} m^a m^b , \quad\quad & \Phi_{12} =
-\frac{1}{2} R_{ab} n^a m^b , & \quad\quad \Phi_{22} = -\frac{1}{2}
R_{ab} n^a n^b , \nonumber \\ & \Lambda = \frac{1}{24} R.
\label{Ricci_np} \end{eqnarray}

\noindent Given the tetrad vectors in Eqs.~\ref{l}-\ref{m}, one can
show that the tetrad components of the Ricci tensor are independent
of $\zeta$, with $\Phi_{00} = \Phi_{22} = 2 \Phi_{11} = (1/2)
\nabla^2 \varphi$ as the only non-zero components.  The tetrad
components of the Weyl tensor are not independent of $\zeta$, but to
zeroth order in $\zeta$ are given in Eq.~\ref{weyl}.

The NP spin coefficients are named, complex combinations of the
Ricci rotation coefficients.  They are defined by

\begin{eqnarray}
\rho = \ell_{a;b}m^a\bar m^b, \quad\quad \sigma = \ell_{a;b}m^a m^b,
& \quad\quad & \kappa = \ell_{a;b}m^a \ell^b, \quad\quad \tau =
\ell_{a;b}m^a n^b, \nonumber \\ \mu = -n_{a;b} \bar m^a m^b ,
\quad\quad \lambda = -n_{a;b} \bar m^a \bar m^b , & \quad\quad & \nu
= -n_{a;b} \bar m^a n^b , \quad\quad \pi = -n_{a;b} \bar m^a \ell^b
, \nonumber \\ \epsilon = \frac{1}{2} \left(\ell_{a;b} n^a \ell^b -
m_{a;b} \bar m^a \ell^b \right) , & \quad\quad & \alpha =
\frac{1}{2} \left(\ell_{a;b} n^a \bar m^b - m_{a;b} \bar m^a \bar
m^b \right), \nonumber \\ \gamma = -\frac{1}{2} \left(n_{a;b} \ell^a
n^b - \bar m_{a;b} m^a n^b \right) , & \quad\quad & \beta =
-\frac{1}{2} \left(n_{a;b} \ell^a m^b - \bar m_{a;b} m^a m^b
\right). \label{spin} \end{eqnarray}

\noindent We are interested in the spin coefficients for the flat
background metric.  To compute the spin coefficients, it is simplest
to change coordinates to $(t,r,\zeta,\bar\zeta)$ where the flat
metric takes the form

\begin{equation} ds^2 = dt^2 - dr^2 - \frac{4 \, r^2 \, d\zeta\,
d\bar\zeta}{(1+\zeta\bar\zeta)^2}, \label{mfapp}  \end{equation}

\noindent and the null tetrad is given by

\begin{eqnarray} \ell^a = \left( \frac{-1}{\sqrt{2}},
\frac{1}{\sqrt{2}},0,0\right), &\quad\quad&  n^a = \left(
\frac{-1}{\sqrt{2}},
\frac{-1}{\sqrt{2}},0,0\right) \nonumber \\
m^a = \left( 0, 0, \frac{1+\zeta\bar\zeta}{\sqrt{2} \, r} ,0\right),
&\quad\quad& \bar m^a = \left( 0, 0, 0,
\frac{1+\zeta\bar\zeta}{\sqrt{2} \, r} \right). \nonumber
\end{eqnarray}

\noindent Although the connection for the flat metric,
Eq.~\ref{mfapp}, is not zero, then zeros in the components of the
tetrad vectors in the $(t,r,\zeta, \bar\zeta)$ coordinate system
make computing the spin coefficients much simpler.  The non-zero
spin coefficients are

\begin{eqnarray} \rho = \mu &=& - \frac{1}{\sqrt{2}r} ,
\label{rho:app} \\ \alpha = \frac{\zeta}{2\,\sqrt{2}\, r} & ~ &
\beta
 \frac{-\bar \zeta}{2\, \sqrt{2} \, r}. \label{beta}
\end{eqnarray}

\noindent Only $\rho$ and $\mu$ are zeroth order in $\zeta$.

%_________________________________________________________

% FIGURES

\end{document}